# Investigation of factors regarding the effects of COVID-19 pandemic on college students' depression by quantum annealer

Junggu Choi[1], Kion Kim[2], Soohyun Park[3], Juyoen Hur[3], Hyunjung Yang[4], Younghoon Kim[3], Hakbae Lee[5,6] & Sanghoon Han[1,3,*]

[1] Yonsei Graduate program in Cognitive Science, Yonsei University, Seoul, 03722, Republic of Korea
[2] Korea Quantum Computing©, Seoul, 06164, Republic of Korea
[3] Department of Psychology, Yonsei University, Seoul, 03722, Republic of Korea
[4] University College, Yonsei University, Seoul, 03722, Republic of Korea
[5] Deparment of Applied Statistics, Yonsei University, Seoul, 03722, Republic of Korea
[6] Department of Statistics and Data science, Yonsei University, Seoul, 03722, Republic of Korea

[*] Corresponding author: sanghoon.han@yonsei.ac.kr

**Abstract**

Diverse cases regarding the impact, with its related factors, of the COVID-19 pandemic on mental health have been reported in previous studies. College student groups have been frequently selected as the target population in previous studies because they are easily affected by pandemics. In this study, multivariable datasets were collected from 751 college students based on the complex relationships between various mental health factors. We utilized quantum annealing (QA)-based feature selection algorithms that were executed by commercial D-Wave quantum computers to determine the changes in the relative importance of the associated factors before and after the pandemic. Multivariable linear regression (MLR) and XGBoost models were also applied to validate the QA-based algorithms. Based on the experimental results, we confirm that QA-based algorithms have comparable capabilities in factor analysis research to the MLR models that have been widely used in previous studies. Furthermore, the performance of the QA-based algorithms was validated through the important factor results from the algorithms. Pandemic-related factors (e.g., confidence in the social system) and psychological factors (e.g., decision-making in uncertain situations) were more important in post-pandemic conditions. We believe that our study will serve as a reference for researchers studying similar topics.

**Introduction**

The coronavirus disease 2019 (COVID-19) pandemic has significantly disrupted global society in diverse aspects [1,2]. Researchers have attempted to investigate the influence of pandemics on various domains, including biological or psychological areas in previous studies [3-5]. Among the several domains impacted by the pandemic, mental health issues during the pandemic received attention as they were associated with regulations, including social distancing [6,7]. In particular, researchers have focused on college students, who can be influenced more easily than other age groups, to investigate differences or variations in their mental health status during the pandemic. For example, several researchers have emphasized the importance of long-term monitoring and mental health support for college students, with experimental results of the analyses of the determinants of mental health [8,9].

Several studies have emphasized post-pandemic depression as a significant mental health challenge [10-12]. The word "Corona-Blue" has been used to represent depression and lethargy due to self-isolation and social distancing in associated research [13,14]. Moreover, the possible determinants of depression were investigated to identify the factors affected by the pandemic. The prevalent social and economic factors that were established as major factors before the pandemic were found to be identical after the pandemic [15,16]. Furthermore, pandemic-related factors, including fear of job loss and the lockdown, were confirmed through statistical validations [17,18].

Based on the various factors related to depression, the collected dataset comprised multiple variables in diverse categories. To identify latent relationships within multivariable datasets, statistical modeling or factor analysis methods have been utilized in previous studies [19-21]. For example, statistical modeling, including multivariable linear regression (MLR) models, has been used to interpret the associations of variables with their respective coefficients [22,23]. In addition, exploratory and confirmatory factor analysis methods have been widely applied to examine symptom dimensions (e.g., insomnia and atypical symptoms) in a large community-based cohort [24-26]. These factor analysis methods need to set prior information (e.g., application of the number of factors or hypotheses by related theories) [27,28].

However, data-driven methods, including machine learning (ML) or deep learning (DL) algorithms without specific prior criteria have been widely used in recent studies. Notably, ML algorithm performances were mainly used as evaluation criteria to validate the impact of diverse factor candidates [29,30]. Additionally, feature importance results, which can be calculated through the working of ML algorithms (e.g., coefficients of support vector machine or f-score from XGBoost algorithms), are utilized to compare the relative effects between variables in datasets [31,32]. Furthermore, several DL algorithms have proposed specialized modules in their process to show important feature sets from input datasets. To detect depression using diverse categories of datasets, self-attention or attention mechanism modules have been applied to analyze the features noticed by algorithms [33,34].

Unlike the two methods (statistical and data-driven methods) mentioned above, a quantum annealing (QA)-based optimization method has been introduced for feature selection in existing studies [35,36]. These methods entail the optimization of energy state (from the initial state to the lowest energy state) in commercial quantum annealers of D-Wave devices. For the selection of various biosignal features to detect stress level, an automated feature selection framework with quantum annealer was proposed [37]. In addition, optimization algorithms using quantum annealing have been applied to select the optimized transcription factors for DNA sequence [38].

In this study, we investigated the changes in the relative importance of variables before and after the pandemic using a survey dataset collected from college student groups on depression. Among the three aforementioned methods that could be applied for factor investigation, we focused on the QA-based feature selection algorithm to validate the algorithm performance through comparisons with MLR models (statistical method). Moreover, XGBoost models (data-driven method) were utilized to evaluate the selected features from the two algorithms (i.e., QA-based algorithms and statistical analysis methods) using their classification and regression performances as quantitative indices. Finally, we confirm the variables with their rank changes after the pandemic using trends proposed in previous studies as a qualitative evaluation tool.

**Results**

*XGBoost algorithm performance of selected features from two feature selection algorithms.* To evaluate the selected feature sets from the D-Wave QA algorithms and MLR models, we checked the performance of the XGBoost algorithms under the classification and regression tasks. Moreover, experiments with five evaluation conditions based on accumulated independent variables (i.e., top 1 ~ 10, top 1 ~ 20, top 1 ~ 30, top 1 ~ 40, and top 1 ~ 50 variables) were conducted to compare the influence of the selected features on the performance of the XGBoost algorithm. All experimental conditions were repeated 30 times to validate the differences between the conditions using t-test.

We found that the overall performance indices (classification: balanced accuracy; regression: negative mean squared error) of the QA-based algorithms were higher than those of the MLR models. Moreover, the differences in the indices between the two methods (QA-based algorithms and MLR models) in all the experimental results were confirmed to be statistically significant (*p-value* < 0.05). The detailed experimental results are presented in Tables 1 and 2.

Table 1. Average classification performance (balanced accuracy) of the XGBoost algorithm

| Experimental conditions | Top 10 (top 1 ~ 10) | Top 20 (top 1 ~ 20) | Top 30 (top 1 ~ 30) | Top 40 (top 1 ~ 40) | Top 50 (top 1 ~ 50) |
|---|---|---|---|---|---|
| Dependent variable | Before the COVID-19 pandemic | | | | |
| Quantum annealer | 0.745 | 0.743 | 0.736 | 0.741 | 0.746 |
| Multivariable linear regression | 0.589 | 0.604 | 0.614 | 0.620 | 0.621 |
| Dependent variable | After the COVID-19 pandemic | | | | |
| Quantum annealer | 0.750 | 0.725 | 0.735 | 0.743 | 0.748 |
| Multivariable linear regression | 0.580 | 0.605 | 0.635 | 0.629 | 0.629 |

The differences in all the averaged performance indices (QA vs. MLR) were statistically significant ($p\text{-}value$ < 0.05).

Table 2. Average regression performance (negative mean absolute error) of the XGBoost algorithm

| Experimental conditions | Top 10 (top 1 ~ 10) | Top 20 (top 1 ~ 20) | Top 30 (top 1 ~ 30) | Top 40 (top 1 ~ 40) | Top 50 (top 1 ~ 50) |
|---|---|---|---|---|---|
| Dependent variable | Before the COVID-19 pandemic | | | | |
| Quantum annealer | -3.152 | -2.950 | -2.870 | -2.819 | -2.829 |
| Multivariable linear regression | -4.531 | -4.152 | -4.051 | -3.989 | -3.948 |
| Dependent variable | After the COVID-19 pandemic | | | | |
| Quantum annealer | -2.959 | -2.959 | -2.819 | -2.820 | -2.798 |
| Multivariable linear regression | -4.386 | -4.168 | -3.988 | -3.843 | -3.852 |

The differences in all the averaged performance indices (QA vs. MLR) were statistically significant ($p\text{-}value$ < 0.05).

Based on comparisons of XGBoost algorithm performances, we identify that QA-based feature selection algorithms have comparable feature selection capabilities to MLR models, which have been widely utilized in previous studies.

Based on the aforementioned trends, we checked the variables with higher ranks in the post-pandemic conditions calculated using QA-based algorithms. In the top 1 to 10 ranks, the ranks of pandemic- and virus-related variables were higher after the pandemic than before the pandemic. For example, the importance of social interactions with family, friends, and third parties ("q226" and "q143") and reliability of the public medical system ("q300") became high.

Similarly, in the top 11 to 50, pandemic (or virus)-related variables and decision-making-associated variables showed higher ranks in post-pandemic conditions. Social distancing ("q303"), safety of family members in pandemic ("q299"), infections about contact with foreigners ("q163"), and exaggerations of COVID-19 dangers ("q159") showed relatively higher importance than before pandemic conditions. In addition, uncertain situations ("q77"), deferred decision-making ("q186"), and deferred decision-making with anxious thoughts ("q134") were also found to have higher ranks in the post-pandemic conditions. The detailed changes in the variables and survey questions for each confirmed variable are presented in Tables 3 and 4.

Table 3. Changes in variable ranks before and after the COVID-19 pandemic from QA-based feature selection algorithms (top 1 ~ top 50).

| Rank | Before | After | Rank | Before | After | Rank | Before | After | Rank | Before | After | Rank | Before | After |
|---|---|---|---|---|---|---|---|---|---|---|---|---|---|---|
| 1 | q_319 | q_319 | 11 | q331 | q302 | 21 | q77 | **q186** | 31 | q73 | q260 | 41 | q298 | q262 |
| 2 | q2_1 | q2_1 | 12 | q330 | **q264** | 22 | q265 | q265 | 32 | q159 | q148 | 42 | q157 | q164 |
| 3 | q317 | q317 | 13 | q266 | q266 | 23 | q186 | q149 | 33 | q255 | q71 | 43 | q329 | **q174** |
| 4 | q318 | q318 | 14 | q143 | q330 | 24 | q258 | **q259** | 34 | q163 | q165 | 44 | q267 | **q161** |
| 5 | q263 | **q226** | 15 | q327 | q332 | 25 | q262 | **q261** | 35 | q261 | **q157** | 45 | q173 | **q181** |
| 6 | q226 | **q143** | 16 | q148 | q301 | 26 | q71 | **q163** | 36 | q162 | **q267** | 46 | q174 | q75 |
| 7 | q302 | **q327** | 17 | q264 | **q303** | 27 | q165 | q258 | 37 | q181 | **q72** | 47 | q168 | q76 |
| 8 | q332 | **q300** | 18 | q300 | q328 | 28 | q260 | **q255** | 38 | q252 | q251 | 48 | q161 | q179 |
| 9 | q328 | **q331** | 19 | q149 | **q77** | 29 | q259 | **q184** | 39 | q75 | q73 | 49 | q250 | q298 |
| 10 | q301 | q263 | 20 | q303 | **q299** | 30 | q184 | **q159** | 40 | q299 | q162 | 50 | q72 | q256 |

Variables with higher rankings in after pandemic conditions are highlighted in bold.

Table 4. Survey questions of highlighted variables with higher ranks in the post-COVID-19 pandemic condition.

| No. | Rank of variables | Survey question code | Survey question |
|---|---|---|---|
| 1 | 5 | q226 | Who has helped you during the COVID-19 pandemic period? |
| 2 | 6 | q143 | Since the COVID-19 pandemic began, how many meetings have you attended with family, friends, or third parties? |
| 3 | 7 | q327 | I can imagine what my life will be like in 10 years. |
| 4 | 8 | q300 | I'm afraid the health care system won't protect my loved ones. |
| 5 | 9 | q331 | I am always hopeful about my future. |
| 6 | 12 | q264 | I prefer work that is important and intellectually challenging to work that is important but does not require much thought. |
| 7 | 17 | q303 | I worry that social distancing won't be enough to keep me safe from the virus. |
| 8 | 19 | q77 | I have to get away from all uncertain situations. |
| 9 | 20 | q299 | I'm worried that I won't be able to keep my family safe from the virus. |
| 10 | 21 | q186 | I defer decision-making whenever possible. |
| 11 | 24 | q259 | I find it fascinating to rely on thinking to reach the top. |
| 12 | 25 | q261 | Learning new ways of thinking doesn't excite me. |
| 13 | 26 | q163 | I am concerned about contact with foreigners as foreigners may be infected with the virus. |
| 14 | 28 | q255 | I like to contemplate for a long time. |
| 15 | 29 | q184 | The thought of having to make a decision makes me anxious, so I keep putting it off. |
| 16 | 30 | q159 | The dangers of COVID-19 have been greatly exaggerated. |
| 17 | 35 | q157 | The government should open the area I live in without shutting it down due to COVID-19. |
| 18 | 36 | q267 | I often find myself thinking about matters that do not affect me personally. |
| 19 | 37 | q72 | I have to be able to plan everything in advance. |
| 20 | 43 | q174 | I tend to rely on intuition when making decisions. |
| 21 | 44 | q161 | I am concerned that foreigners are spreading the virus in our country. |
| 22 | 45 | q181 | I often make decisions impulsively. |

## Methods

*Study design.* To compare the different factors before and after the COVID-19 pandemic, our study adopted five steps. First, we collected survey datasets from 751 college students using 560 survey questions in 14 categories, including demographics, pre-pandemic, and post-graduation. Second, the collected survey datasets were preprocessed by removing variables with missing values and checking for possible outliers. Third, the preprocessed datasets were applied to quantum annealer and MLR models to determine the relative importance of the variables from each algorithm. Fourth, the variables from the two algorithms (quantum annealer and MLR models) were validated using XGBoost algorithms with regression and classification performance. Finally, the rank change of the variables from the algorithm that exhibited better performance was confirmed. Our overall study design is depicted in Fig. 1.

Figure 1. Overview of the study design

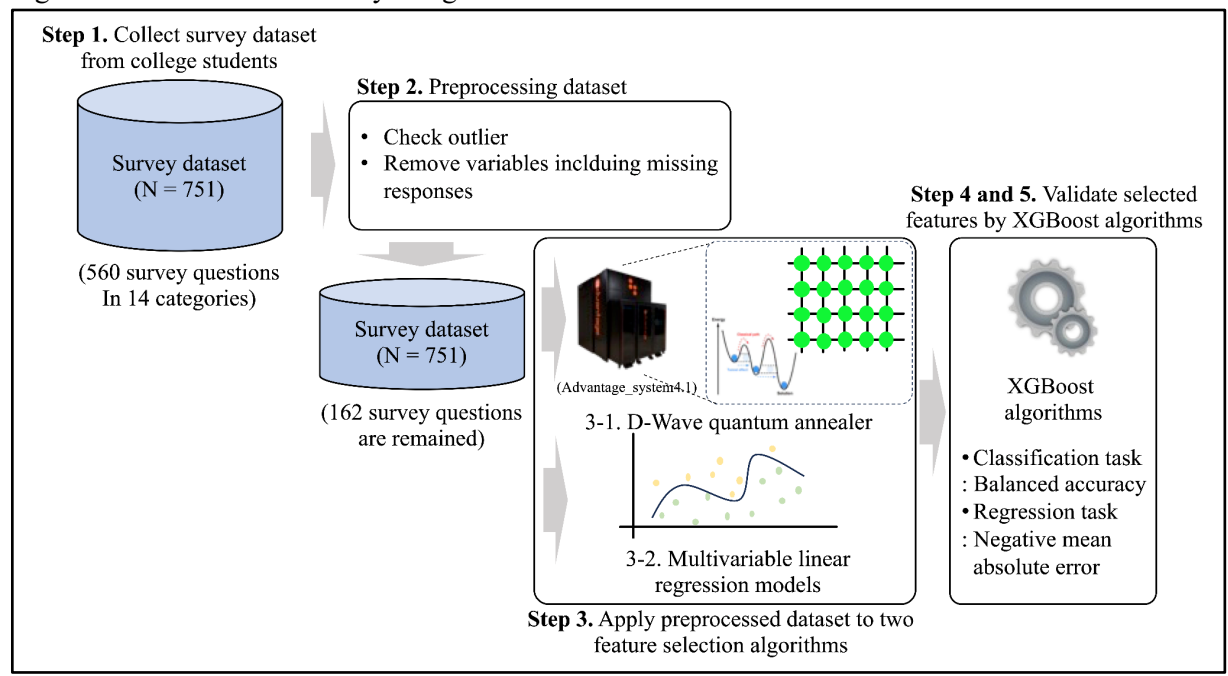

*Datasets and preprocessing.* We collected responses to 560 survey questions in 14 categories from 751 college students (Male: 220, Female: 531, Age: 22.15 $\pm$ 2.15). To investigate the mental health of college students during the pandemic, associated variables—anxiety, depression, suicide, panic, and traumatic events—were added to the survey. In addition, pandemic-related survey categories—social distancing and coping with the pandemic—were added. The raw survey dataset was (751, 560) (i.e., 751 rows and 560 columns).

Before applying the algorithms for variable importance, we removed variables (columns) with missing values. After removing columns with missing values, the shape of the dataset changed to (751, 270) (i.e., 290 variables were removed). Among the diverse variables related to mental health, 18 depression-related variables were set as the dependent variables. Because mental health variables constituted variables about "About 2 weeks before the COVID-19 pandemic" (i.e., before pandemic) and variables about "In the past 2 weeks" (after pandemic), nine variables each for before and after the pandemic were summed to single variables. Furthermore, all variables with similar content were excluded. The final dataset comprised 751 rows and 162 columns (variables).

*D-Wave quantum processor unit (QPU).* A D-Wave QPU was built based on the Ising model using statistical mechanics. This model comprised lattice structures with nodes and edges [39]. In each node, the spins are located as binary variables. In a D-Wave processor, superconducting flux qubits represent the spins. The variables in the Ising model indicate the spin-up and spin-down states corresponding to +1 and -1 values, respectively. The correlations between the spins were calculated to represent their relationships. The objective function of the Ising model is as follows:

$$E_{ising}(S) = \sum_{i=1}^{N} h_i s_i + \sum_{i=1}^{N} \sum_{j=i+1}^{N} J_{i,j} s_i s_j \quad (1)$$

where $h_i$ represents the bias of the $i^{th}$ qubit and $s_i$ indicates spin in the spin set ($s_i \in S$) of the quantum system. In addition, $J_{i,j}$ is the coupling strength between $i^{th}$ spin and $j^{th}$ spin, and $E_{ising}(S)$ denotes the Hamiltonian of the quantum system in a given state $S$. The optimization process aims to minimize the energy state to determine the system's ground state.

One can leverage the Ising model to solve diverse optimization problems by translating it into a binary quadratic model (BQM) [40]. The objective function of the BQM is as follows:

$$f(x) = \sum_{i} Q_{i,i} x_i + \sum_{i<j} Q_{i,j} x_i x_j = x^T Q x \quad (2)$$

where $Q_{i,j}$ is the element in the $i^{th}$ row and $j^{th}$ column of a real-valued upper-triangular matrix $Q$ and $x_i \in \{0,1\}$ is the $i^{th}$ element of $x$. The objective function $f(x)$ is minimized to find optimal solutions (i.e., $argmin_{x \in \{0,1\}^n} x^T Q x$) for a given coefficient matrix $Q$.

*Mapping features to D-Wave quantum annealer.* We used mutual information (MI) to reflect the correlation between features and conditional information (CMI) to incorporate the relationships between features and dependent variables. The CMI values should be negated during the minimization process since we wanted to maximize them. MI and CMI are defined as follows:

$$I(X;Y) = H(X) - H(X|Y) \quad (3)$$

$$I(X;Y|Z) = H(X|Z) - H(X|Y,Z) \quad (4)$$

where $I(X;Y)$ and $I(X;Y|Z)$ represent MI and CMI, respectively. In Equations (3) and (4), $H(X)$ denotes the marginal entropy of $X$, quantifying the amount of information contained in the features, and $H(X|Y)$ indicates the conditional entropy. We used negative CMIs as the diagonal terms and MIs as the upper-triangular terms of the coefficient matrix $Q$.

The BQM library included in the D-Wave Ocean software converts the objective function to a graph $G = (V, E)$, where $V$ denotes a set of vertices, and $E$ denotes a set of edges connecting vertices that encode MIs and CMIs and embeds it into D-Wave's QPU. To select the best feature combination, the BQM underwent an iterative optimization process for a given number of features, $k$, which ranges from 1 to 161. After completing the entire process, the relative importance of the features was obtained by counting the number of appearances. For more details regarding graph-level implementation, see [41].

*Statistical models for comparison.* To compare the feature selection results from the D-Wave QA based algorithms, we used MLR models. The same features and dependent variables were used in the MLR models. The coefficient values of each feature from the fitted MLR models were set as the criteria for relative associations with the dependent variables.

*XGBoost algorithms for the validation of selected feature sets.* After organizing the two selected feature sets calculated from the two algorithms (D-Wave QA-based algorithms and MLR models), we need to validate the feature selection results. Among the various applicable methods, XGBoost algorithms, which were widely known in classical ML algorithms, were selected as an evaluation model. The classification and regression performance index values from the trained XGBoost algorithms were compared for the two feature selection algorithms. The dependent variables were additionally converted from continuous to binary categorical variables for evaluation in the classification task of the XGBoost algorithms. Among the 162 selected variables with their ranks, we conducted accumulative evaluations with 5 variable conditions based on the rank of each variable (top 10, top 20, top 30, top 40, and top 50). Balanced accuracy and negative mean absolute error were utilized as performance indices for each classification and regression task.

*Tools.* D-Wave QA-based feature selection algorithm codes were written using the D-Wave Ocean software (Python-based software). Data preprocessing, MLR models, and XGBoost classifiers were built and operationalized using Python (version 3.7.1; scikit-learn, version 2.4.1) and R (version 4.0.3).

**Discussion**

In this study, we investigated the factors affecting depression in college student groups by comparing depression-related factors during the pre-pandemic period. To compare the appropriacy of diverse feature selection algorithms, we considered three types of algorithms in our study design (i.e., statistical modeling, data-driven modeling, and QA-based algorithms). Among the three algorithms considered, we focused on appraising the feature selection capability of the QA-based algorithms through comparisons with statistical modeling (MLR models). Feature sets with importance ranks calculated from the QA-based algorithms and MLR models were validated using data-driven methods (XGBoost algorithms under classification and regression tasks) to evaluate each feature selection algorithm.

Various feature selection methodologies have been utilized to identify the major factors in the datasets in previous studies [42,43]. To examine the applicability of QA-based algorithms for feature selection tasks, we applied MLR models, which could be considered as typical methods utilized in associated studies [44]. In addition, XGBoost algorithms are widely used in ML models for multivariate analysis as validation models to evaluate selected features based on their importance, according to algorithm performances [45]. Two performance indices (balanced accuracy and negative mean-squared error) were used for the classification and regression tasks.

For validation experiments with XGBoost algorithms, we created experimental conditions with accumulated variables based on the rank of each variable. Five conditions were fixed with groups including 10 variables (i.e., condition 1: top 1 ~ 10, condition 2: top 1 ~ 20, condition 3: top 1 ~ 30, condition 4: top 1 ~ 40, and condition 5: top 1 ~ 50). We hypothesize that if we could check a similar scale of performance indices in the overall conditions, it indicates that the 50 features selected from each algorithm have sufficient associations with depression in college student groups (dependent variables). Moreover, if higher performance indices are established in QA-based algorithms, it indicates that such QA-based algorithms can show capabilities similar to those of existing methods that have been applied in previous studies.

In the classification tasks with selected features, we verified that the QA-based algorithms showed higher performance (higher balanced accuracy) in all experimental conditions. Similarly, in the regression tasks, smaller negative mean absolute errors were ascertained using the QA-based algorithms in all experimental conditions. From the aforementioned results, we confirmed that the QA-based feature selection algorithm is competitive compared with the MLR models. Furthermore, performance index values showed similar trends from condition 1 (top 1 ~ 10 variables) to condition 5 (top 1 ~ top 50), and these were found in most experimental conditions except the "after pandemic" condition in classification tasks. Balanced accuracy in the classification tasks increased and negative mean absolute error in the regression tasks decreased from condition 1 to condition 5 (i.e., balanced accuracy in condition 5 was higher than that in condition 1, and negative mean absolute error in condition 5 was lower than that in condition 1). Based on these trends, we also checked the feature selection capacity of the QA-based algorithms.

Based on the quantitative evaluation mentioned above, we confirm that the QA-based feature selection algorithms are applicable to feature selection tasks based on comparisons with the methods utilized in previous studies. Based on these results, we compared the important features before and after the pandemic using QA-based algorithms.

Among the diverse variables that needed to be validated, COVID-19-related variables showed more significant changes with higher ranks after pandemic than before it. Especially, variables about social interactions in pandemic were checked in the top-10 category (e.g., "q226": "Who has helped you during the COVID-19 pandemic period?" and "q143": "Since the COVID-19 pandemic began, how many meetings have you attended with family, friends, or third parties?"). Similar trends regarding the impact of social interaction during pandemics on depression have been reported in previous studies [46, 47]. In addition to the depression of older adult groups during the pandemic, research on the influence of depression on young adults and adolescent groups during the pandemic was conducted [48, 49]. Constraints such as the COVID-19 home confinement and school concerns were found to be the main factors vis-à-vis depression among young adults and adolescents [50]. Moreover, the importance of social relationship replacements (e.g., social media usage) was checked for depression and loneliness in the young adult group [51].

Additionally, social systems, including public medical system-related variables, showed increased ranks after the pandemic. Variables about the concern about the public medical system and safety of family were checked (e.g., "q300": "I'm afraid the healthcare system won't protect my loved ones" and "q299": "I'm worried that I won't be able to keep my family safe from the virus"). Furthermore, social-constraint-related variables also showed higher ranks in after pandemic ("q303": "I worry that social distancing won't be enough to keep me safe from the virus" and "q157": "The government should open the area I live in without shutting it down due to COVID-19"). Concerns about social or medical systems have been investigated as factors related to depression in young adults in associated studies [52, 53]. Relationships between social distancing, including quarantine, with psychosocial consequences, and well-being have been confirmed in young adult groups [54]. Unlike external elements in a pandemic (e.g., social distancing or social systems), internal elements, including psychological factors, were examined together with higher ranks in our experimental results. For example, hope about the future and leaving from uncertain situations were found ("q331": "I am always hopeful about my future" and "q77": "I have to get away from all uncertain situations").

In summary, we verified that QA-based algorithms can be used for feature selection in multivariable datasets, based on comparisons with MLR models (quantitative evaluations). Moreover, the selected variables related to depression in college student groups (i.e., young adult groups) by QA-based algorithms included several variables in both the external and internal element categories (qualitative evaluations). These trends in the selected variables were validated with previous studies.

## Conclusion

In this study, we analyzed the important variables related to depression in college student groups before and after the COVID-19 pandemic using QA-based feature selection algorithms. The QA-based algorithms, executed on the D-Wave QPU, are validated by comparisons with MLR and XGBoost algorithms, which have been widely utilized in previous studies. Through validation experiments, we identify that QA-based algorithms have a feature selection capability that is as good as that of previously applied methods. Social interactions and social systems in the pandemic-related variables ranked higher after the pandemic. Moreover, psychological factor variables, including decision-making in uncertain situations and hope for the future, ranked higher after the pandemic. These results were additionally verified with previous studies.

Our study has several strengths from diverse perspectives. First, we applied QA-based feature selection algorithms using D-Wave hardware to investigate the survey datasets collected from psychological studies. Second, our experimental results verified the performance of the QA-based algorithm in our research settings. Third, our research can propose usage examples of D-Wave QA for researchers working on COVID-19-pandemic-associated research topics. However, our study has some limitations. First, to generalize our research conclusions, an analysis of datasets collected from other countries should be conducted using the same research scheme. Second, we need to analyze datasets for other mental-health-related variables (e.g., anxiety or loneliness) to identify complex variables for the overall mental health status of college students.

## Ethics approval
This study was approved by the Institutional Review Board (IRB) Committee of Yonsei University (7001988-2021110-HR-1244-05), and all tests were carried out in accordance with relevant guidelines and regulations.

## Consent to participate
Informed consent was obtained from all individual participants included in the study.

## Consent to publish
The authors affirm that the human research participants provided informed consent for publication.

## Funding
This work was supported in part by the Yonsei Signature Research Cluster Program of 2021, under grant number 2021-22-0005 and in part by the quantum computing technology development program of the National Research Foundation of Korea(NRF) funded by the Korean government (Ministry of Science and ICT(MSIT)) (No. 2020M3H3A1110365).

## Acknowledgments
We would like to thank the Quantum Information Research Support Center at Sungkyunkwan University that provided support regarding D-Wave quantum computing resources for this research.

## Author contributions
J.G.C: formal analysis, visualization, writing - original draft, review and editing; K.I.K: formal analysis, writing - original draft, review and editing; S.H.P: data collection, investigation; J.Y.H: data collection, investigation; H.J.Y: data collection, investigation; Y.H.K: data collection, investigation; H.B.L: formal analysis, writing - review and editing; S.H.H: formal analysis, writing - review and editing, supervision

**Competing interests**

The authors declare no competing interests.

**References**

1. Chakraborty, I., & Maity, P. COVID-19 outbreak: Migration, effects on society, global environment and prevention. *Science of the total environment.* **728**, 138882 (2020).
2. Onyeaka, H. *et al.* COVID-19 pandemic: A review of the global lockdown and its far-reaching effects. *Science progress* **2**, 104 (2021).
3. De Figueiredo, C. S. *et al.* COVID-19 pandemic impact on children and adolescents' mental health: Biological, environmental, and social factors. *Progress in Neuro-Psychopharmacology and Biological Psychiatry* **106**, 110171 (2021).
4. Gildner, T. E., & Thayer, Z. M. Maternal and child health during the COVID-19 pandemic: Contributions in the field of human biology. *American Journal of Human Biology.* **32**, 5 (2020).
5. O'regan, D., Jackson, M. L., Young, A. H., & Rosenzweig, I. Understanding the impact of the COVID-19 pandemic, lockdowns and social isolation on sleep quality. *Nature and science of sleep.* **13**, 2053-2064 2021.
6. Carvalho Aguiar Melo, M., & de Sousa Soares, D. Impact of social distancing on mental health during the COVID-19 pandemic: An urgent discussion. *International Journal of Social Psychiatry.* **66**, 625-626 (2020).
7. Banks, J., & Xu, X. The mental health effects of the first two months of lockdown during the COVID-19 pandemic in the UK. *Fiscal Studies.* **41**, 685-708 (2020).
8. Chen, T., & Lucock, M. The mental health of university students during the COVID-19 pandemic: An online survey in the UK. *PloS one.* **17**, e0262562 (2022).
9. Wang, X. *et al.* Investigating mental health of US college students during the COVID-19 pandemic: Cross-sectional survey study. *Journal of medical Internet research.* **22**, e22817 (2020).
10. Ustun, G. Determining depression and related factors in a society affected by COVID-19 pandemic. *International Journal of Social Psychiatry.* **67**, 54-63 (2021).
11. Salari, N. *et al.* Prevalence of stress, anxiety, depression among the general population during the COVID-19 pandemic: a systematic review and meta-analysis. *Globalization and health.* **16**, 1-11 (2020).
12. Ettman, C. K. *et al.* Prevalence of depression symptoms in US adults before and during the COVID-19 pandemic. *JAMA network open.* **3**, e2019686-e2019686 (2020).
13. Myung Suk, A. A public perception study on the new word "Corona Blue": focusing on social media big data analysis. *International Journal of Advanced Culture Technology (IJACT).* **8**, 133-139 (2020).
14. Lee, J. *et al.* Risk perception, unhealthy behavior, and anxiety due to viral epidemic among healthcare workers: the relationships with depressive and insomnia symptoms during COVID-19. *Frontiers in Psychiatry.* **12**, 358 (2021).
15. Langer, Á. I. *et al.* Social and economic factors associated with subthreshold and major depressive episode in university students during the Covid-19 pandemic. *Frontiers in public health.* **10**, 893483 (2022).
16. Rondung, E., Leiler, A., Meurling, J., & Bjärtå, A. (2021). Symptoms of depression and anxiety during the early phase of the COVID-19 pandemic in Sweden. *Frontiers in Public Health*, *9*, 562437.
17. Khubchandani, J. *et al.* Post-lockdown depression and anxiety in the USA during the COVID-19 pandemic. *Journal of Public Health.* **43**, 246-253 (2021).
18. Benke, C., Autenrieth, L. K., Asselmann, E., & Pané-Farré, C. A. Lockdown, quarantine measures, and social distancing: Associations with depression, anxiety and distress at the beginning of the COVID-19 pandemic among adults from Germany. *Psychiatry research.* **293**, 113462 (2020).
19. Krieger, T. *et al.* Self-compassion in depression: Associations with depressive symptoms, rumination, and avoidance in depressed outpatients. *Behavior therapy.* **44**, 501-513 (2013).
20. Dudek, D. *et al.* Risk factors of treatment resistance in major depression: association with bipolarity. *Journal of affective disorders.* **126**, 268-271 (2010).
21. Norton, S. *et al.* The Hospital Anxiety and Depression Scale: a meta confirmatory factor analysis. *Journal of psychosomatic research.* **74**, 74-81 (2013).
22. Kong, X. *et al.* Prevalence and factors associated with depression and anxiety of hospitalized patients with COVID-19. Preprint at https://www.medrxiv.org/content/10.1101/2020.03.24.20043075v2 (2020).

23. Karaçam, Z., & Ançel, G. Depression, anxiety and influencing factors in pregnancy: a study in a Turkish population. *Midwifery*. **25**, 344-356 (2009).
24. Wainberg, M. *et al.* Symptom dimensions of major depression in a large community-based cohort. *Psychological Medicine*. **53**, 438-445 (2023).
25. Clara, I. P., Cox, B. J., & Enns, M. W. Confirmatory factor analysis of the Depression–Anxiety–Stress Scales in depressed and anxious patients. *Journal of psychopathology and behavioral assessment*. **23**, 61-67 (2001).
26. Smith, A. B. *et al.* Factor analysis of the Hospital Anxiety and Depression Scale from a large cancer population. *Psychology and Psychotherapy: Theory, Research and Practice*. **75**, 165-176 (2002).
27. Dunbar, M., Ford, G., Hunt, K., & Der, G. A confirmatory factor analysis of the Hospital Anxiety and Depression scale: comparing empirically and theoretically derived structures. *British Journal of Clinical Psychology*. **39**, 79-94 (2000).
28. Ollendick, T. H. *et al.* Anxiety and depression in children and adolescents: A factor-analytic examination of the tripartite model. *Journal of Child and Family Studies*. **12**, 157-170 (2003).
29. Xu, W. *et al.* Risk factors analysis of COVID-19 patients with ARDS and prediction based on machine learning. *Scientific reports*. **11**, 2933 (2021).
30. Qasrawi, R. *et al.* Assessment and prediction of depression and anxiety risk factors in schoolchildren: machine learning techniques performance analysis. *JMIR Formative Research*. **6**, e32736 (2022).
31. Oh, T. *et al.* Machine learning-based diagnosis and risk factor analysis of cardiocerebrovascular disease based on KNHANES. *Scientific Reports*. **12**, 2250 (2022).
32. Zhang, W. *et al.* Machine learning models for the prediction of postpartum depression: application and comparison based on a cohort study. *JMIR medical informatics*. **8**, e15516 (2020).
33. Du, Z., Li, W., Huang, D., & Wang, Y. Encoding visual behaviors with attentive temporal convolution for depression prediction. *IEEE international conference on automatic face & gesture recognition*, 1-7 (2019).
34. Saggu, G. S., Gupta, K., Arya, K. V., & Rodriguez, C. R. DepressNet: A Multimodal Hierarchical Attention Mechanism approach for Depression Detection. *Int. J. Eng. Sci.* **15**, 24-32 (2022).
35. Otgonbaatar, S., & Datcu, M. A quantum annealer for subset feature selection and the classification of hyperspectral images. *IEEE Journal of Selected Topics in Applied Earth Observations and Remote Sensing*. **14**, 7057-7065 (2021).
36. Ferrari Dacrema, M. *et al.* Towards feature selection for ranking and classification exploiting quantum annealers. *Proceedings of the 45th International ACM SIGIR Conference on Research and Development in Information Retrieval*. 2814-2824 (2022).
37. Nath, R. K., Thapliyal, H., & Humble, T. S. Quantum annealing for automated feature selection in stress detection. *IEEE Computer Society Annual Symposium on VLSI (ISVLSI)*. 453-457 (2021).
38. Li, R. Y., Di Felice, R., Rohs, R., & Lidar, D. A. Quantum annealing versus classical machine learning applied to a simplified computational biology problem. *NPJ quantum information*. **4**, 14 (2018).
39. Cipra, B. A. An introduction to the Ising model. *The American Mathematical Monthly*. **94**, 937-959 (1987).
40. Glover, F., Kochenberger, G., & Du, Y. A tutorial on formulating and using QUBO models. Preprint at https://arxiv.org/abs/1811.11538 (2018).
41. Harris, R. *et al.* Experimental demonstration of a robust and scalable flux qubit. *Physical Review B*. **81.13**, 134510 (2010).
42. Ballard, E. D. *et al.* Parsing the heterogeneity of depression: An exploratory factor analysis across commonly used depression rating scales. *Journal of Affective Disorders*. **231**, 51-57 (2018).
43. Khodayari-Rostamabad, A. *et al.* A machine learning approach using EEG data to predict response to SSRI treatment for major depressive disorder. *Clinical Neurophysiology*. **124**, 1975-1985 (2013).
44. Musil, C. M., Jones, S. L., & Warner, C. D. Structural equation modeling and its relationship to multiple regression and factor analysis. *Research in Nursing & Health*. **21**, 271-281 (1998).
45. Sharma, A., & Verbeke, W. J. Improving diagnosis of depression with XGBOOST machine learning model and a large biomarkers Dutch dataset (n= 11,081). *Frontiers in big Data*. **3**, 15 (2020).
46. Noguchi, T. *et al.* Living Alone and Depressive Symptoms among Older Adults in the COVID-19 Pandemic: Role of Non–Face-to-Face Social Interactions. Journal of the American Medical Directors Association. **24**, 17-21 (2023).

47. Towner, E. *et al.* Virtual Social Interaction and Loneliness Among Emerging Adults Amid the COVID-19 Pandemic. *Current Research in Ecological and Social Psychology*, **3**, 100058 (2021).
48. Lee, C. M., Cadigan, J. M., & Rhew, I. C. Increases in loneliness among young adults during the COVID-19 pandemic and association with increases in mental health problems. *Journal of Adolescent Health.* **67**, 714-717 (2020).
49. Tüzün, Z., Başar, K., & Akgül, S. Social connectedness matters: Depression and anxiety in transgender youth during the COVID-19 pandemic. *The journal of sexual medicine.* **19**, 650-660 (2022).
50. Hawes, M. T. *et al.* Increases in depression and anxiety symptoms in adolescents and young adults during the COVID-19 pandemic. Psychological medicine. **52**, 3222-3230 (2022).
51. Lisitsa, E. *et al.* Loneliness among young adults during COVID-19 pandemic: The mediational roles of social media use and social support seeking. *Journal of Social and Clinical Psychology.* **39**, 708-726 (2020).
52. Uribe-Restrepo, J. M. *et al.* Mental health and psychosocial impact of the COVID-19 pandemic and social distancing measures among young adults in Bogotá, Colombia. *AIMS Public Health.* **9**, 630 (2022).
53. Benke, C., Autenrieth, L. K., Asselmann, E., & Pané-Farré, C. A. Lockdown, quarantine measures, and social distancing: Associations with depression, anxiety and distress at the beginning of the COVID-19 pandemic among adults from Germany. *Psychiatry research.* **293**, 113462 (2020).
54. Rodríguez-Fernández, P. *et al.* Exploring the occupational balance of young adults during social distancing measures in the COVID-19 pandemic. *International journal of environmental research and public health.* **18**, 5809 (2021).